\documentstyle[aps,epsf,eqsecnum,prb,tighten]{revtex}

\begin{document}
\title{The Kondo Screening Cloud}
\author{Ian Affleck\thanks{on leave from Canadian Institute for Advanced
Research and Physics Department, University of British Columbia}}
\address{Physics Department, Boston University, 590 Commonwealth Ave. \\
Boston, MA02215}
\maketitle
\abstract{Renormalization group theory of the Kondo effect predicts 
that an impurity spin is screened by a conduction electron spread 
over a large distance of order .1 to 1 micron.  This review has the 
following sections:\\
1. The Kondo effect and the screening cloud\\
2. Non-observation of the Kondo cloud in conventional experiments\\
3. Kondo effect in transmission through a quantum dot\\
4. Observing the screening cloud in persistent current experiments\\
5. Side-coupled quantum dot\\
6. Conclusions
}

\section{The Kondo effect and the screening cloud}
\begin{figure}
\vglue 0cm
\hspace{0.01\hsize}
\epsfxsize=1.0\hsize
\epsffile{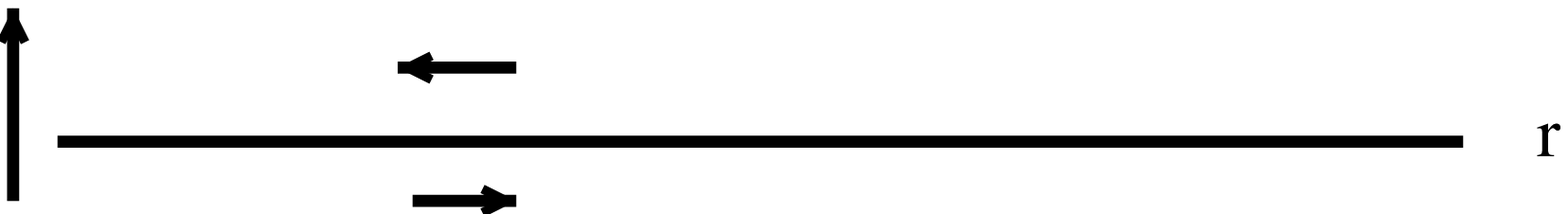}
\label{fig:1D}
\caption{}
\end{figure}

A single impurity in a metal is described by the Kondo (or s-d) model:
\begin{equation}
H=\sum_{\vec k\sigma}\psi^\dagger_{\vec k\sigma}\psi_{\vec k\sigma}
\epsilon_k + J\vec S_{\hbox{imp}}\cdot \vec S_{\hbox{el}}(r=0).
\end{equation}
Here $\vec S_{\hbox{imp}}$ is the impurity spin operator (with S=1/2)
and $\vec S_{\hbox{el}}$ is the electron spin density at position $\vec r$.
After expanding the electron field, $\psi (\vec r)$, in spherical 
harmonics and keeping only the s-wave and linearizing the dispersion 
relation we obtain a relativistic quantum field theory, defined on a half-line
with the impurity at the origin \cite{Affleck}.  (See fig. \ref{fig:1D}.)

The Hamiltonian reduces to:
\begin{equation}
H=iv_F\int_0^\infty dx\left[ \psi^\dagger_L{d\over dx}\psi_L
-\psi^\dagger_R{d\over dx}\psi_R\right] +2\pi v_F\lambda 
\vec S_{\hbox{imp}}\cdot \vec S_{\hbox{el}}(0).\end{equation}
Here $\lambda$ is the dimensionless Kondo coupling constant, $J\nu$, 
where $\nu$ is the density of states.  To study the problem at low 
energies, we may apply the renormalization group, integrating out 
high energy Fourier modes of the electron operators, reducing the 
band-with, $D$:
\begin{equation}
d\lambda /d\ln D \approx -\lambda^2+\ldots \end{equation}
with solution:
\begin{equation}
\lambda_{\hbox{eff}}(D) \approx {\lambda_0\over 1-\lambda_0\ln (D_0/D)}
+\ldots \end{equation}
The effective coupling becomes O(1) at an energy scale $T_K$:
\begin{equation}
T_K\approx De^{-1/\lambda_0}.\end{equation}
Here $\lambda_0$ is the bare Kondo coupling and $D_0$ is of order 
the Fermi energy.  After reducing the bandwidth the effective Hamiltonian 
has a {\it wave-vector} cut off:
\begin{equation}
|k-k_F|<T_K/v_F\equiv \xi_K.\end{equation}
This defines a characteristic {\it length scale} for the Kondo effect; 
it is typically around .1 to 1 micron.  

\begin{figure}
\begin{center}
\vglue 0cm
\hspace{0.01\hsize}
\epsfxsize=.5\hsize
\epsffile{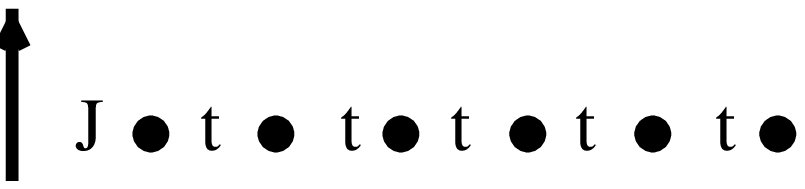}
\end{center}
\caption{}
\label{fig:1Dt}
\end{figure}

At low energies, $T<<T_K$, $\lambda_{\hbox{eff}}$ seems to get 
large.  This strong coupling physics is easiest to understand in 
a tight-binding model. 
\begin{equation}
H=-t\sum_{j=0}^\infty (\psi^\dagger_j\psi_{j+1}+h.c.)+JS_{\hbox{imp}}
\cdot \vec S_{\hbox{el}}(0).\end{equation}
(See fig. \ref{fig:1Dt}.)
For $J>>t$, we simply find the groundstate of the last term in this 
Hamiltonian.  This has 1 electron at site $j=0$ forming a singlet 
with the impurity, 
$|\phi_0>=(|\uparrow \downarrow >-|\downarrow \uparrow >)/\sqrt{2}$. 
The other electrons are free except that they must not go to $j=0$ 
since they would break the singlet.  Effectively they feel an infinite 
repulsion at $j=0$, corresponding to a $\pi /2$ phase shift.  For 
finite (small) $\lambda_0$, this description only holds at low energies 
and small $|k-k_F|$.  Only long wavelength probes see this simple 
$\pi /2$ phase shift.  This is the basis of Nozi\`eres' 
local Fermi liquid theory of the Kondo effect \cite{Nozieres}.
 The short distance 
physics is more complicated, involving the singlet formation.  
Heuristically, we may think of an electron in a wave-function 
which is spread out over this large distance, $\xi_K$, which 
is forming a singlet with the impurity.  (More accurately, we 
should think of the singlet as being formed by a linear 
superposition of an electron and a hole since there is not 
neccessarily any local modification of the charge density around 
the impurity.)
\begin{figure}
\begin{center}
\vglue 0cm
\hspace{0.01\hsize}
\epsfxsize=.5\hsize
\epsffile{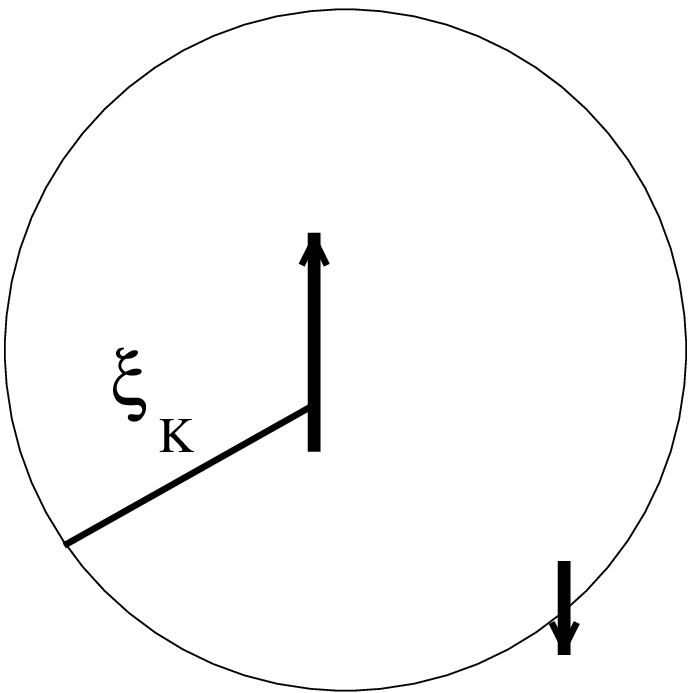}
\caption{}
\end{center}
\label{fig:cloud}
\end{figure}

\section{Non-observation of the screening cloud in conventional Kondo 
experiments}
This long length scale is surprisingly difficult to observe, perhaps 
mainly because it so large.  

An important point to realise is that 
single impurity Kondo behaviour can be observed even when the average 
inter-impurity separation, $R_{\hbox{imp}}$, is much greater than $\xi_K$.
 This happens because the screening cloud wave-functions 
 from different impurities 
are nearly orthogonal even when they are strongly overlapping in 
space \cite{Barzykin}.  
The condition for single impurity Kondo behavior is probably 
\begin{equation}
R_{\hbox{imp}}>>\xi_K^{1/3}k_F^{-2/3}.\end{equation}
This follows from Nozi\`eres ``exhaustion principle'' which states 
that the number of electrons within $|k-k_F|<\xi_K^{-1}$ should be 
 greater than the number of impurities.  This, perhaps optimistic, 
estimate assumes that if enough states are available the individual 
screening clouds will manage to be almost orthogonal.  However, 
in a one-dimensional 
system, the condition is simply that $R_{\hbox{imp}}>>\xi_K$.  

The Knight shift as a function of distance from an impurity can 
be measured by nuclear magnetic resonance \cite{Boyce}.   At $T=0$ this takes 
the form \cite{Barzykin2}:
\begin{equation}
\chi (r)=\chi_0+{\cos (2k_Fr)\over r^2}f\left({r\over \xi_K}\right),
\end{equation}
where $\chi_0$ is the susceptibility of the pure system and $f$ is 
 some universal scaling function.  The rapid oscillations 
and the power law pre-factor make this scaling behavior difficult to 
observe.  Typically the signal gets far too small long before 
$r$ is as large as $\xi_K$.  Furthermore, at such long distances 
the Knight shift will be given by a superposition of contributions 
from many impurities.

The charge density around a Kondo impurity varies only over short 
length scales of $O(1/k_F)$.  For instance, in a particle-hole 
symmetric model, such as a tight-binding model at half-filling, 
the charge density can easily be proven to be completely uniform.  
This is connected with ``spin-charge separation'' in this effectively 
one-dimensional problem.  The Kondo effect takes place purely in 
the spin sector.
Furthermore, the energy and $r$-dependent density of states, probed 
by scanning tunnelling microscopy, only varies on short length 
scales \cite{Ujsaghy}.
For the case of a $\delta$-function Kondo interaction, the electron 
self-energy can be easily shown to have the form:
\begin{equation}
\Sigma (r,\omega ) \propto G_0(r,\omega ) T(\omega )
G_0(r,\omega ).\end{equation}
Here  $G_0(r,\omega )$ is the free electron Green's function, which has 
trivial dependence on $r$.  $T(\omega )$ has non-trivial dependence 
on $\omega$ and varies on the characteristic energy scale $T_K$.  However, 
the $r$-dependence of $\Sigma$, and hence the density of 
states, {\it is} trivial.  

\begin{figure}
\vglue 0cm
\hspace{0.01\hsize}
\epsfxsize=1.0\hsize
\epsffile{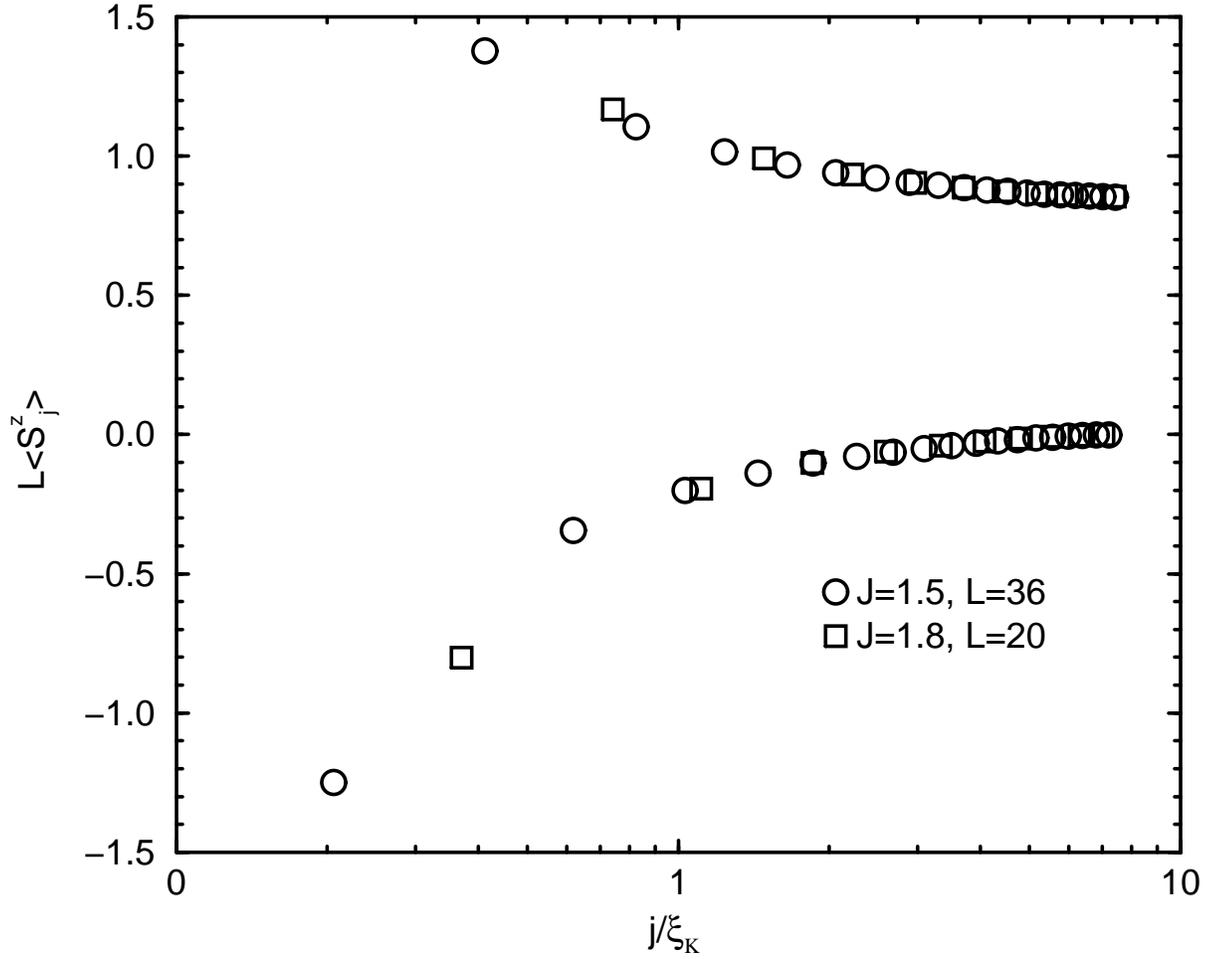}
\label{fig:Szj}
\caption{$L$ times the expectation value of the z-component of the
electron spin, $<S^z_j>$, as a function of $j/\xi_K(J)$.
Two systems are shown: $J=1.8,\ \xi_K=2.7,\ L=20$ and
$J=1.5,\ \xi_K=4.85,\ L=36$
Thus in both cases we have $L/\xi_K\approx 7.4$.
Clearly the data collapses onto
a universal curve.}
\end{figure}

\begin{figure}
\vglue 0cm
\hspace{0.01\hsize}
\epsfxsize=1.0\hsize
\epsffile{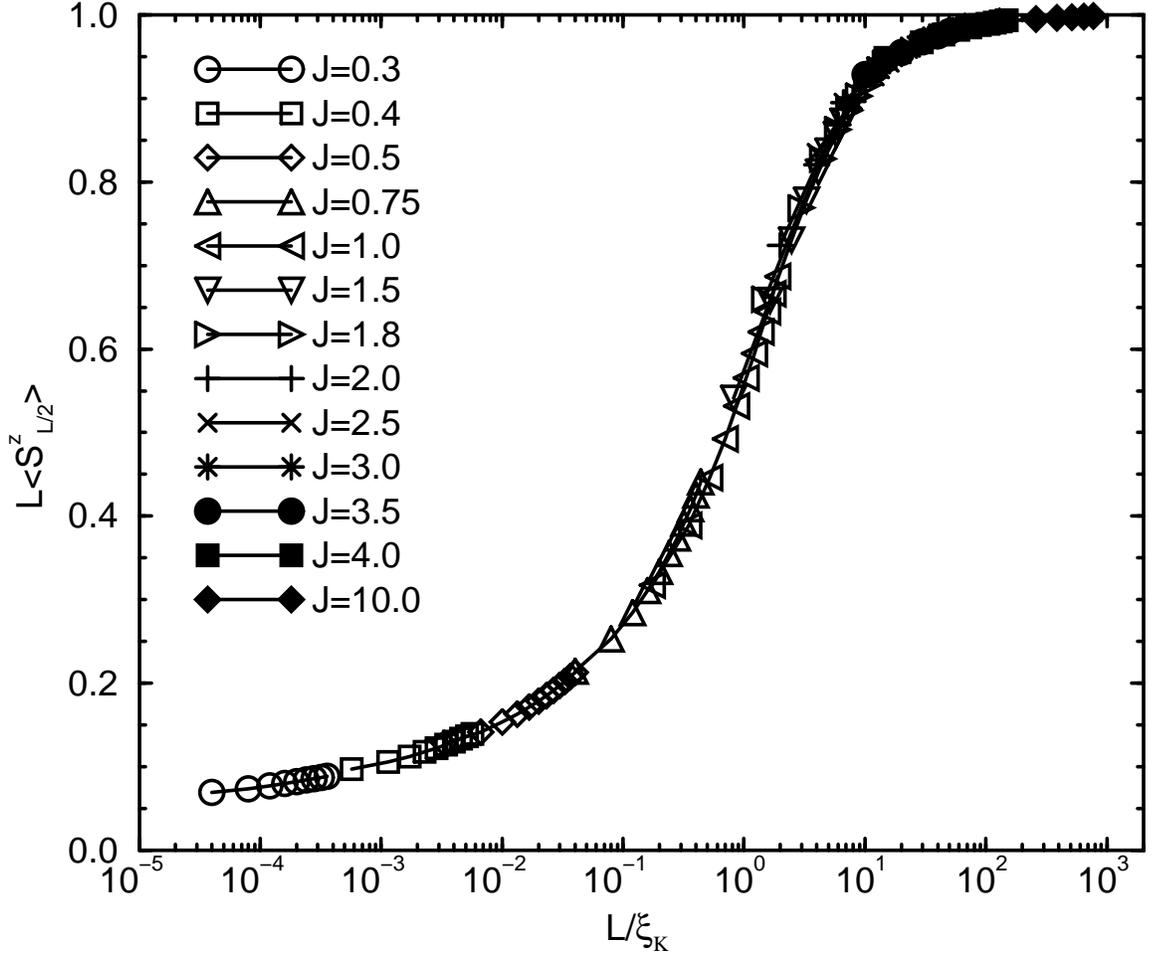}
\label{fig:SzL}
\caption{ Logaritmic plot of $L<S^z_{L/2}>$
as a function of chain length $L/\xi_K$ for
a range of different coupling constants.
The initial point corresponds in
all cases to $L=4$.
The solid lines are guides to the eye.
The strong coupling limit corresponds to $L<S^z_{L/2}>\approx 1$}
\end{figure}

This stubborn refusal of the Kondo length scale to show up in 
experiments might make one wonder if it really exists.  E. S\o rensen 
and I demonstrated \cite{Sorensen} that it does exist by doing density matrix 
renormalization group simulations on large finite chains.  The 
system we studied is the one sketched in fig. \ref{fig:1Dt}, with 
an even number of sites $L$ and open boundary conditions and the
electron density set at 1/2-filling.  In this case the groundstate 
has spin S=1/2.  We measured $<S^z_j>$ in the groundstate 
with $S^z_T=+1/2$.  This takes the value corresponding to 
a non-interacting chain with 1 site excised far from the screening 
cloud ($j>>\xi_K$).  This is $O(1/L)$ and oscillates.  At shorter distances 
$j\leq \xi_K$ $<S^z_j>$ exhibits more complicated behavior.  We showed 
that the data appears to collapse onto a single scaling curve when 
plotted vs. $j/\xi_K$. (See figs. \ref{fig:Szj} and \ref{fig:SzL}.)
  The values of $\xi_K$ obtained from 
this data collapse had the expected exponential dependence on $J$. 

The question remains whether such behaviour could ever be
observed experimentally.  

\section{Kondo effect in transmission through a quantum dot}
\begin{figure}
\begin{center}
\vglue 0cm
\hspace{0.01\hsize}
\epsfxsize=.5\hsize
\epsffile{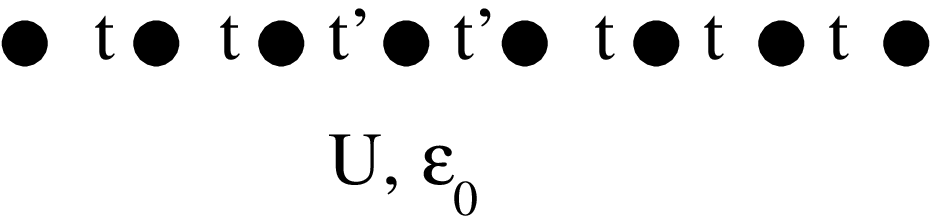}
\caption{}
\end{center}
\label{fig:anderson}
\end{figure}

Quantum dots and wires can be made by first forming a two-dimensional 
electron gas in a semi-conductor heterostructure layer and then 
further confining the electrons by etching and gate voltages varying 
at the sub-micron  scale. Typical dots have been formed with radius 
around .1 micron containing around 50 electrons.  The number of 
electrons on the dot can be varied in unit steps by varying a gate voltage. 
Such a system is a single electron transistor.  When the tunnelling 
amplitude between the dots and the leads is weak such a system 
can exhibit the ``Coulomb blockade''.  The energy cost to add (or 
remove) an electron to the dot becomes sigificant and inhibits 
conductance through the dot at low $T$.  When the number of 
electrons on the dot is {\it odd} it acts like an $S=1/2$ impurity.  
Now a type of Kondo effect takes place which corresponds to {\it perfect 
transmission} rather than perfect reflection 
\cite{Goldhaber,Cronenwett,Simmel}.
  A simplified model 
which captures the essential physics is the one-dimensional Anderson 
model \cite{Glazman,Ng}.  See fig. \ref{fig:anderson}.
\begin{equation}
H=-t\left[\sum_{j\leq -2}+\sum_{j\geq 1}\right](\psi^\dagger_j\psi_{j+1}
+ h.c.) -t'[\psi^\dagger_0(\psi_{-1}+\psi_1)+h.c.] +\epsilon_0\psi^\dagger 
\psi_0+Un_{0\uparrow}n_{0\downarrow}.\label{Anderson}\end{equation}
The conductance is determined by the transmission amplitude through the 
dot at the Fermi energy.  In the non-interacting case ($U=0$) there is 
a transmission resonance when $\epsilon_0=2(t'^2-t^2)\cos k_F$, with 
a width of $O(t')$.  In the Kondo limit, $t'<<-\epsilon_0, U+\epsilon_0$, 
the dot acts like an S=1/2 impurity with a Kondo interaction:
\begin{equation}
H_{\hbox{int}}=J(\psi^\dagger_{-1}+\psi^\dagger_1){\vec \sigma \over 2}
(\psi_{-1}+\psi_1)\cdot \vec S_{\hbox{imp}}.\label{large-J}
\label{Jdot}\end{equation}
Here $\vec S_{\hbox{imp}}$ represents the electron spin on the dot 
and 
\begin{equation}
J=2t'^2[-1/\epsilon_0+1/(U+\epsilon_0)].\end{equation}
As usual, we expect $J$ to renormalize to large values as the band-width 
is reduced.  Again it is useful to consider the behaviour for 
large bare coupling, $J$.  Now the screening electron goes into the symmetric 
orbital on sites 1 and (-1).  Electrons are transmitted through the dot, 
in this limit, by passing through the anti-symmetric orbital:
\begin{equation}
\psi_a\equiv (\psi_1-\psi_{-1})/\sqrt{2}.\end{equation}
The effective low energy Hamiltonian is obtained by taking $J\to \infty$ 
and projecting out the symmetric orbital that screens the impurity.  
\begin{equation}
H--t\left[\sum_{j\leq -2}+\sum_{j\geq 1}\right](\psi^\dagger_j\psi_{j+1}
+ h.c.) -(t/\sqrt{2})
[-\psi^\dagger_{-2}\psi_{a}+\psi^\dagger_a\psi_2++h.c.].\end{equation}
This non-interacting Hamiltonian exhibits perfect conductance at
half-filling.
  
In general additional potential scattering terms are 
induced in the Kondo Hamiltonian of Eq. (\ref{Jdot}) 
which are also of $O(J)$.  This generally occurs unless the 
Hamiltonian has particle-hole symmetry.  (For the Anderson model of 
Eq. (\ref{Anderson}), exact particle-hole symmetry requires half-filling 
and $\epsilon_0=0$.) 
If these 
were added to the large-$J$ effective Hamiltonian they would in 
general move it off resonance and reduce the conductance.  However, 
another important result about the Kondo model is that such particle-hole 
symmetry breaking is strictly marginal.  These terms do not grow 
large at low energies but remain of order the bare Kondo coupling.  Thus 
for small bare Kondo coupling the resonance remains pinned at essentially 
the Fermi energy.  Consequently there is a plateau in the transmission 
as a function of gate voltage, $\epsilon_0$, over which the conductance 
is close to the idea value $2e^2/h$, with a width of $O(U)$.  Thus, 
near perfect conductance occurs at low $T$ over the entire range 
of gate voltage where the occupancy of the dot is an odd integer.    
Such low temperature conductance plateaux have been observed by 
the Delft group \cite{Wiel}.
\section{Observing the screening cloud in a persistent current 
experiment}
\begin{figure}
\begin{center}
\vglue 0cm
\hspace{0.01\hsize}
\epsfxsize=.5\hsize
\epsffile{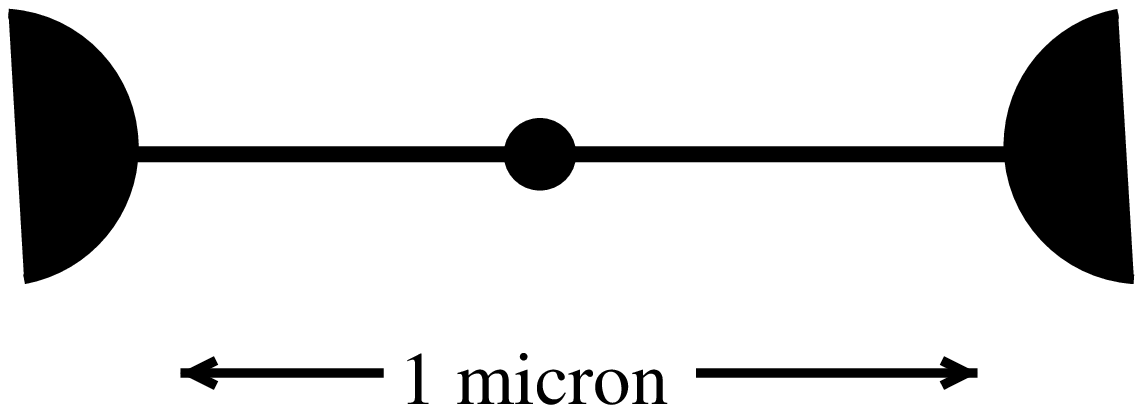}
\caption{}
\end{center}
\label{fig:qwire}
\end{figure}

It is convenient to consider a narrow quantum wire connected to 
the quantum dot which is eventually connected to macroscopic leads. 
(See fig. \ref{fig:qwire}.)
Now the screening could will live in the  leads.  
It is likely to be considerably larger than the quantum dot and 
perhaps about equal to the length of the quantum wire leads 
in some experiments (eg. those of the Delft group).  However, 
these quantum wires are generally connected to macroscopic leads, 
as sketched in fig. \ref{fig:qwire}.  The screening cloud 
can also exist in the macroscopic section of the leads.  Thus it 
is not obvious how the size of the screening cloud will manifest 
itself in most experimental set-ups.  

\begin{figure}
\begin{center}
\vglue 0cm
\hspace{0.01\hsize}
\epsfxsize=.25\hsize
\epsffile{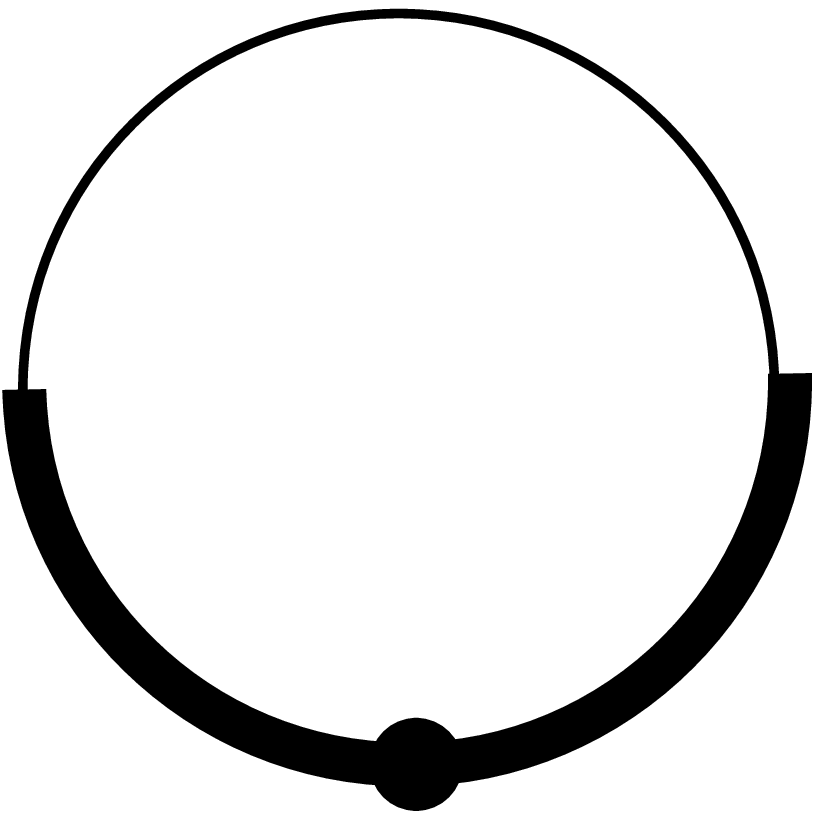}
\caption{}
\end{center}
\label{fig:ring}
\end{figure}

One simple possibility (from a theoretical point of view) is to 
study a closed ring containing a quantum dot.  (See fig. \ref{fig:ring}.)
Now the screening cloud is trapped on the ring and can't escape into 
any macroscopic leads.  The experimental difficulty is to measure 
the current through the dot!  This could perhaps be done by applying 
a magnetic field to the ring and measuring the resulting persistent 
current.  When the cloud size, $\xi_K<<L$, the ring circumference, 
the perfect transmission through the dot, due to the Kondo effect, 
implies that the persistent current should be the same as for an 
ideal ring with no dot.  On the other hand, when $\xi_K>>L$, the 
Kondo effect doesn't take place; the infrared divergence of the 
Kondo coupling, $\lambda$, is cut off by the finite size of the ring. 
In this limit the persistent current can be calculated perturbatively 
in $\lambda$. 

\begin{figure}
\begin{center}
\vglue 0cm
\hspace{0.01\hsize}
\epsfxsize=1.0 \hsize
\epsffile{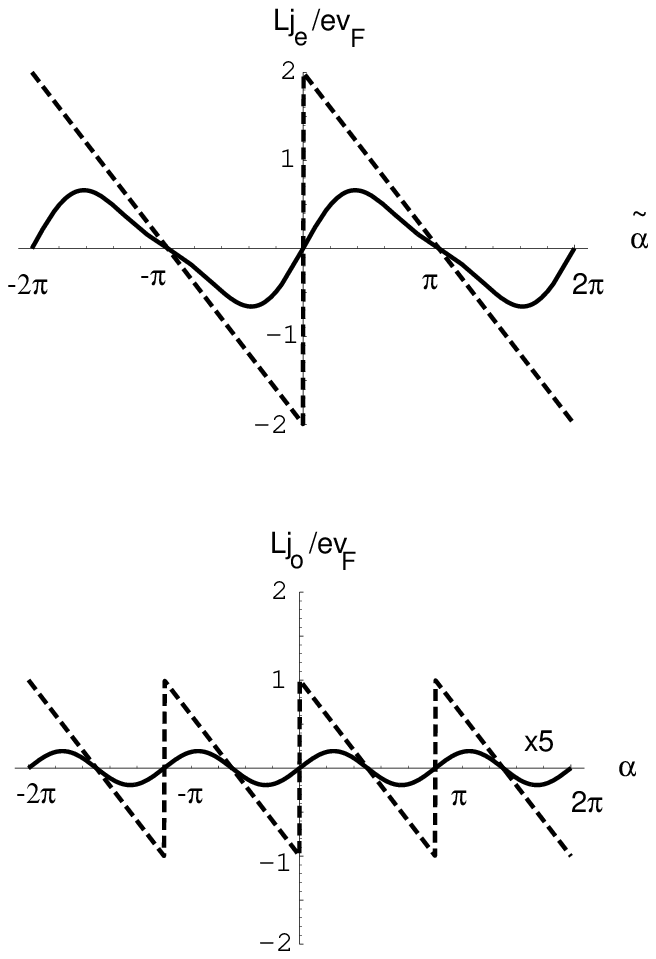}
\caption{ Persistent current vs. flux for an even or odd number of 
electrons for $\xi_K/L\approx 50$ (solid line) and for $\xi_K/L<<1$
 (dashed line).  $j_o$ 
is multiplied $\times 5$ at $\xi_K/L=50$ for visibility.  The 
solid lines are obtained from Eq. (\ref{jpert}) using the effective 
coupling $\lambda (L) \approx 1/\ln (\xi_K/L)$.}
\end{center}
\label{fig:current}
\end{figure}

Analytical expressions for low order perturbation theory can be derived
 in the tight binding model, for 
$L>>1$, by using continuum limit approximations: linearizing the 
dispersion relation, etc \cite{Affleck2}.
  To include magnetic flux we simply give 
the hopping terms to the dot phases $\pm \alpha /2$:
\begin{equation}
H_{\hbox{int}}\to 
J(e^{i\alpha /2}\psi^\dagger_{-1}+e^{-i\alpha /2}\psi^\dagger_1)
{\vec \sigma \over 2}
(e^{-i\alpha /2}\psi_{-1}+e^{i\alpha /2}\psi_1)
\cdot \vec S_{\hbox{imp}}.\label{large-Jflux}
\end{equation}
The corresponding flux is $\Phi = (c/2)\alpha$.  The persistent 
current (at $T=0$) is calculated from the flux-dependence  of the 
groundstate energy: 
\begin{equation}
j=-cdE/d\Phi .\end{equation}
When the total number of electrons (including the electron 
on the dot), $N$, is even, there is 1 unpaired electron at the 
Fermi surface which forms a singlet with the impurity spin.  This 
leads to a contribution of $E$ of first order in $\lambda$.  When 
$N$ is odd, $E$ is second order in $\lambda$ since the groundstate 
of the $(N-1)$ non-interacting electrons is a spin singlet.  To 
calculate next order corrections in $\lambda$ we use the continuum 
propogator:
\begin{equation}
G(\tau ,x) \approx \sum_{n=0}^\infty e^{-(v_F\tau +ix)\pi n/L}=
1/(1-e^{-v_F\tau +ix)\pi /L}).\end{equation}
This leads to the following expressions, for $N$ even or odd:
\begin{eqnarray}
j_e(\alpha )&\approx& {3\pi v_Fe\over 4L}\{\sin \tilde \alpha [
\lambda +\lambda^2 \ln Lc]+(1/4+\ln 2)\lambda^2\sin 2\tilde \alpha \} 
+\ldots \nonumber \\
j_0(\alpha )&\approx &  {3\pi v_Fe\over 16L}\sin 2\alpha [\lambda^2+
2\lambda^3\ln Lc']+\ldots .\label{jpert}\end{eqnarray}
Here $\tilde \alpha$ is $\alpha$ for $N/2$ even or $\alpha +\pi$ for 
$N/2$ odd.  $c$ and $c'$ are dimensionless constants which 
we have not bothered to compute. 
Note that $Lj$ depends on $\lambda$ and $L$ only through 
the renormalized coupling constant at scale $L$:
\begin{equation}
\lambda_{\hbox{eff}}(L)\approx \lambda + \lambda^2\ln L +\cdots .
\end{equation}
This is expected to be exactly true, in all orders of perturbation 
theory due to standard scaling arguments.  Perturbation theory 
breaks down when $\xi_K\leq L$ but we expect $Lj$ to be given 
by some universal scaling functions of $\xi_K/L$.  These need 
to be calculated from large scale numerical simulations, or perhaps 
using integrability methods.  The current in the two limits 
$L<<\xi_K$ and $L>>\xi_K$ are plotted in fig. \ref{fig:current}.  In 
the latter case we simply use the result for an ideal ring with 
no quantum dot.  We expect the universal scaling functions to 
give a smooth cross over between these two limits.  

Our results disagree with those of several other 
groups \cite{Zvyagin,Eckle,Cho,Anda}.

There are many effects left out of this simple model which are probably 
important to a complete understanding of potential experiments in 
this field.  One of these is the presence of several channels in 
the quantum wire.  The Delft experiments suggest about 5 partially filled 
1D bands.  If we also include more structure in the quantum dot then 
it might be appropriate to consider a multi-channel Kondo model 
with all the Kondo couplings, $\lambda_i$,  different.  
The corresponding RG equations, to third order, are:\cite{Blandin}
\begin{equation}
-d\lambda_i/d\ln D = \lambda_i^2-(1/2)\lambda_i\sum_j\lambda_j^2+\ldots 
\end{equation}
For unequal bare couplings, these equations predict that the largest 
one gets larger and the rest shrink towards zero under renormalization.  
The physical picture is that one channel screens the quantum dot 
and has perfect conductance at $T\to 0$, but the other channels 
have zero conductance at $T\to 0$.  Thus the basic picture of 
the Kondo effect is not changed.  Assymetric leads (with 
different hopping parameters $t'_L$ and $t'_R$ on the two sides 
of the quantum dot) reduce the conductance and hence the maximum 
current but we still expect a cross over behaviour at $L\approx \xi_K$. 
Luttinger liquid interactions in the quantum wire destroys the plateau 
in the conductance vs. gate voltage, leaving perfect conductance 
only at one resonant value of the gate voltage \cite{KF} (or none if the 
interactions are strong enough \cite{Kim} ), at least in a one channel model.  
Non-magnetic disorder in the leads reduces the persistent current 
even in the limit $L>>\xi_K$.
\section{Side-coupled dot}
Another interesting model has the quantum dot ``side-coupled'' to 
the  quantum wire ring, as shown in fig. \ref{fig:SC}.  The simplified 
Anderson Hamiltonian now becomes:
\begin{figure}
\begin{center}
\vglue 0cm
\hspace{0.01\hsize}
\epsfxsize.5\hsize
\epsffile{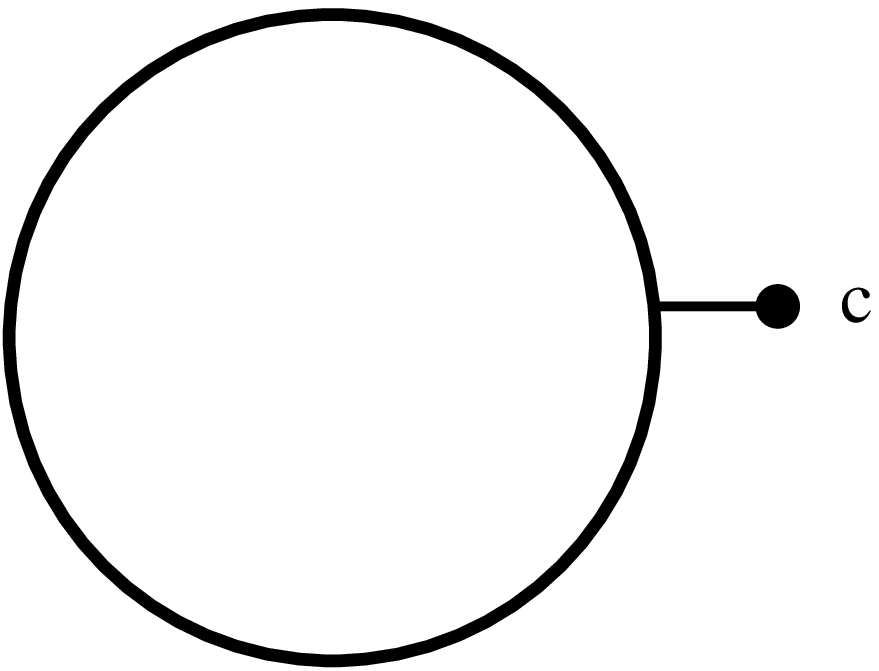}
\caption{}
\end{center}
\label{fig:SC}
\end{figure}
\begin{equation}
H=-t\sum_{j=0}^{L-1}](\psi^\dagger_j\psi_{j+1}
+ h.c.) -t'[\psi^\dagger_0c+h.c.]+ \epsilon_0c^\dagger 
c+Un_{c\uparrow}n_{c\downarrow}.\end{equation}
Now when $t'=0$ there is a perfect current.  The growth of the Kondo 
coupling under renormalization drives the current to 0 for a large ring. 
The formation of the screening cloud now interferes with the current. 
Again this is clear in the large $J$ limit.  A single electron sits 
at site 0 and screens the impurity.  Other electrons must stay away 
from 0 corresponding to a $\pi /2$ phase shift in the even channel.  
The odd channel wave-functions vanish at 0 so there is no conduction 
route for electrons in the large $J$ limit.  
Again we predict scaling behaviour in $\xi_K/L$.  The difference beween 
the two cases may be understood from the formula for the transmission 
through an impurity with even and odd phase shifts:
\begin{equation}
|T|^2=\sin^2(\delta_e-\delta_o),\end{equation}
where $\delta_e$ and $\delta_o$ are the phase shirts at the Fermi 
energy in the even and odd channel.  When the effective even phase shift 
goes from near 0  to near $\pi /2$ for a long ring, $T$ can switch 
between values near 0 and 1.  Which occurs in which limit depends 
on $\delta_0$.  
\section{Conclusions}
The Kondo effect due to spin impurites in metals or due to a quantum 
dot always involves a large screening cloud whenever the dimensionless 
Kondo coupling is small.  This has never been observed 
experimentally.  In mesoscopic experiments the screening cloud 
may ``escape'' into macroscopic leads, in general.  This can be 
avoided in a closed ring experiment where a clear experimental 
signal emerges from the dependence of the persistent current on 
the ratio $\xi_K/L$.  How screening manifests iteself in transmission 
experiment with open leads appears much more subtle 
\cite{Thimm,Cornaglia,Simon2}.


\end{document}